\documentclass[a4paper, 10pt, preprint]{elsarticle}

\usepackage{helvet} 
\usepackage{amsfonts, amsmath, amsthm, amssymb}
\usepackage{graphicx}
\usepackage{booktabs} 
\usepackage[labelfont=bf]{caption} 
\usepackage[top=0.5in,bottom=0.5in,left=0.5in,right=0.5in]{geometry} 
\usepackage{stfloats}
\usepackage{cleveref}

\usepackage{mathtools}
\usepackage{amsmath}
\usepackage{amsmath,esint}
\usepackage{threeparttable}
\usepackage{standalone}
\usepackage{booktabs, dcolumn}
\usepackage{bigints}

\author[ube]{D.~Gamborino\corref{cor1}}

\author[ube]{P.~Wurz}

\cortext[cor1]{Corresponding author \\ Email: diana.gamborino@space.unibe.ch \\ Telephone: +41 0316314427}

\address[ube]{Physikalisches Institut, Universit\"at Bern, Sidlerstrasse 5, CH-3012 Bern, Switzerland}

\journal{Planetary and Space Science}

\begin{document}

\begin{frontmatter}

\title{Velocity distribution function of Na released by photons from planetary surfaces}

\begin{abstract}

In most surface-bound exospheres Na has been observed at altitudes above 
what is possible by thermal release. Photon stimulated desorption of 
adsorbed Na on solid surfaces has been commonly used to explain observations 
at high altitudes. We investigate three model velocity distribution functions 
(VDF) that have been previously used in several studies to describe the 
desorption of atoms from a solid surface either by electron or by photon 
bombardment, namely: the Maxwell-Boltzmann (M-B) distribution, the empirical 
distribution proposed by \cite{Wu10} for PSD, and the Weibull distribution. 
We use all available measurements reported by \cite{Yak00, Yak04} to test 
these distributions and determine which one fits best (statistically) and 
we discuss their physical validity. Our results show that the measured VDF
of released Na atoms are too narrow compared to Maxwell-Boltzmann fits with 
supra-temperatures as suggested by \cite{Yak04}. We found that a good fit 
with M-B is only achieved with a speed offset of the whole distribution to 
higher speeds and a lower temperature, with the offset and the fit temperature 
not showing any correlation with the surface temperature. From the three 
distributions we studied, we find that the Weibull distribution provides the 
best fits using the temperature of the surface, though an offset towards 
higher speeds is required. This work confirms that Electron-Stimulated Desorption 
(ESD) and Photon-Stimulated Desorption (PSD) should produce non-thermal velocity 
(or energy) distributions of the atoms released via these processes, which is 
expected from surface physics. We recommend to use the Weibull distribution 
with the shape parameter $\kappa$=1.7, the speed offset $v_0$ = 575 m/s, and the surface 
temperature to model PSD distributions at planetary bodies.

\end{abstract}

%% ------------------------------------------------------------------------------ %%


\end{frontmatter}

\section{Introduction}

Since the advent of space exploration and ground-based observations that made 
possible the detection of, for instance, the tenuous Na and K atmospheres of Mercury 
and the Moon, several studies and experiments have been carried out on lunar 
mineral grains and simulant materials (e.g.\ \cite{Kell93}). Some studies have 
focused on investigating the most relevant desorption processes for alkalis
(e.g. \cite{Ageev98, Wil99, Yak00, Yak04, Mad98}) to better understand the interaction 
between the surface and the exosphere of the planetary body under study.

Neutral sodium in the exospheres of the Moon and Mercury is one of the most studied 
alkali metals since it is relatively easy to observe from the Earth, but for several 
decades there has been controversy concerning the processes promoting it into the 
atmosphere. One of the few experimental results in the laboratory studying the 
sodium release processes happening on Mercury's surface are the experiments by 
\cite{Yak00, Yak04}. These authors studied the desorption induced by electronic transitions 
(DIET) of Na adsorbed on model mineral surfaces and lunar basalt samples. In 
particular, they measured velocity distribution functions (VDF) of Na released via ESD from SiO$_2$ 
surfaces and found it to be ``clearly non-thermal'' with respect to the surface 
temperature, similar to that of a 1200~K Maxwellian but with a higher-energy tail. 
The VDF of ESD from a lunar basalt sample was found to have a smaller 
offset in speed compared to that of SiO$_2$, with a peak around 0.8~km/s instead of 
1~km/s \cite{Yak04}. Since ESD is a charge transfer process leading to electronic 
excitations similar to PSD, with comparable cross sections and an identical excitation threshold of 
$\geq$4 to 5 eV \cite{Yak00}, the VDF distributions of released sodium 
are quite similar, so that ESD measurements can be substituted for the effects of 
UV photons. Since then, people have interpreted these VDFs using either 
thermal (Maxwellian) or non-thermal distributions. In previous studies a Maxwell-
Boltzmann velocity distribution has been assumed with temperatures in the range: 
$T_s=$ 1200--1500~K (e.g.\ \cite{Sar12, Kill09, Leb03, Kill07}).

In other studies, non-thermal high-energy tail distributions have been 
assumed, for instance the energy distribution function (EDF) used by \cite{John02} for 
ESD from icy surfaces, which was later used and modified by \cite{Wu10} 
to model PSD of volatiles to determine the Na and K density profiles in the exosphere 
of Mercury. This modified version was also used by \cite{Sch12} and 
\cite{Mu09} to determine the escape rates of PSD process in Mercury's 
exosphere, and by \cite{Ten13} and \cite{Spra12} to model lunar Na exosphere.
A summary of previous works using different EDFs and 
arriving at different temperatures of the released Na atoms by PSD or 
ESD is shown in Table~\ref{resumen}.

\begin{table}
\centering
\caption{Summary of different EDFs or VDFs used.}
\begin{tabular}{l l l l}
\hline
\textbf{Reference} & \textbf{Process} & \textbf{EDF/VDF} & \textbf{\it{T} (K)}\\
\hline
\cite{John02}		& ESD	& $CEU^{\beta}/(E+U)^{2+\beta}$ (100 K ice surfaces)  	 	& 600 \\ 
\cite{Leb03}		& PSD	& M--B (fit to \cite{Yak00})						 		& 1500 \\
\cite{Wu10}			& PSD	& adapted from \cite{John02}								& 1500 \\ 
\cite{Sar12}		& PSD	& M--B (model lunar exosphere)			 					& 1200 \\ 
\cite{Sch13}		& PSD	& Kappa (fit to \cite{Yak00})								& 800 \\ 
\cite{Sch13}		& PSD	& M--B (fit to \cite{Yak04})						 		& 500 \\ 
\hline
\end{tabular}
\label{resumen}
\end{table}

Hitherto we use the measurement results reported by \cite{Yak00, Yak04}, 
more specifically, the reported VDFs for neutral Na from SiO$_2$ substrates 
and from lunar basalt samples. We examine the fitness of some distributions 
functions, namely the Maxwell-Boltzmann, the \textit{empirical} energy 
distribution proposed by \cite{Wu10} for released volatiles from Mercury's 
surface via PSD (named here after ``E-PSD'') which is based on the one used 
by \cite{John02} for icy surfaces, and the Weibull distribution. Using the 
Graphical Residual Analysis (GRA), we determine which of the these distributions 
is statistically more adequate to explain the measurements and we discuss their 
physical validity.

The way the energy is imparted to a photodesorbed atom from Mercury's surface (or 
similar planetary surfaces) is not through a thermal process, but rather by single 
electronic excitations. Choosing an appropriate model of the EDF/VDF of the atoms 
released is important to properly interpret Na measurements 
in planetary exospheres, which are often assumed to have temperatures way above the 
surface temperature (see review by \cite{Kill07} or the work by 
\cite{Cass15}, for instance). This work aims to clarify the 
implications of assuming either thermal or non-thermal energy distributions of atoms 
released by PSD and ESD from planetary surfaces not protected by an atmosphere, like 
the majority of the planetary objects of the solar system. 

In Section \ref{psd} we give a general physical description of ESD and PSD processes, 
and in Section \ref{exp} we describe the results from experiments by \cite{Yak00, Yak04}. 
The measurements reported from these experiments are 
used for the statistical analysis in Section \ref{prob}, where we present the 
mathematical description of the different probability distribution functions used 
to fit these measurements. We briefly describe in Section \ref{gra} the GRA we 
used to test the model distribution described in the previous section. In Section 
\ref{res} we show the results of the fitting and the GRA, we discuss the physical 
interpretations in Section \ref{dis}, and we conclude in Section \ref{concl}.

%.................................................................................%%
\section{Desorption Induced by Electronic Transitions (DIET)}
\label{psd}

DIET phenomenon refers to both the electron-stimulated desorption (ESD) and
the photon-stimulated desorption (PSD). Desorption of atoms on the surface occurs 
when the surface is bombarded by electrons or by photons with sufficient energy to 
induce transitions to repulsive electronic states of the atom. The released 
particles are supra-thermal because the absorbed UV photon has energies 
way in excess compared to thermal energies of the surface, which leads to the 
excitation of an anti-bonding state, see Fig. 2 for a schematic representation of 
the process.

\begin{figure}[h]
\centering
\includegraphics[scale=0.22]{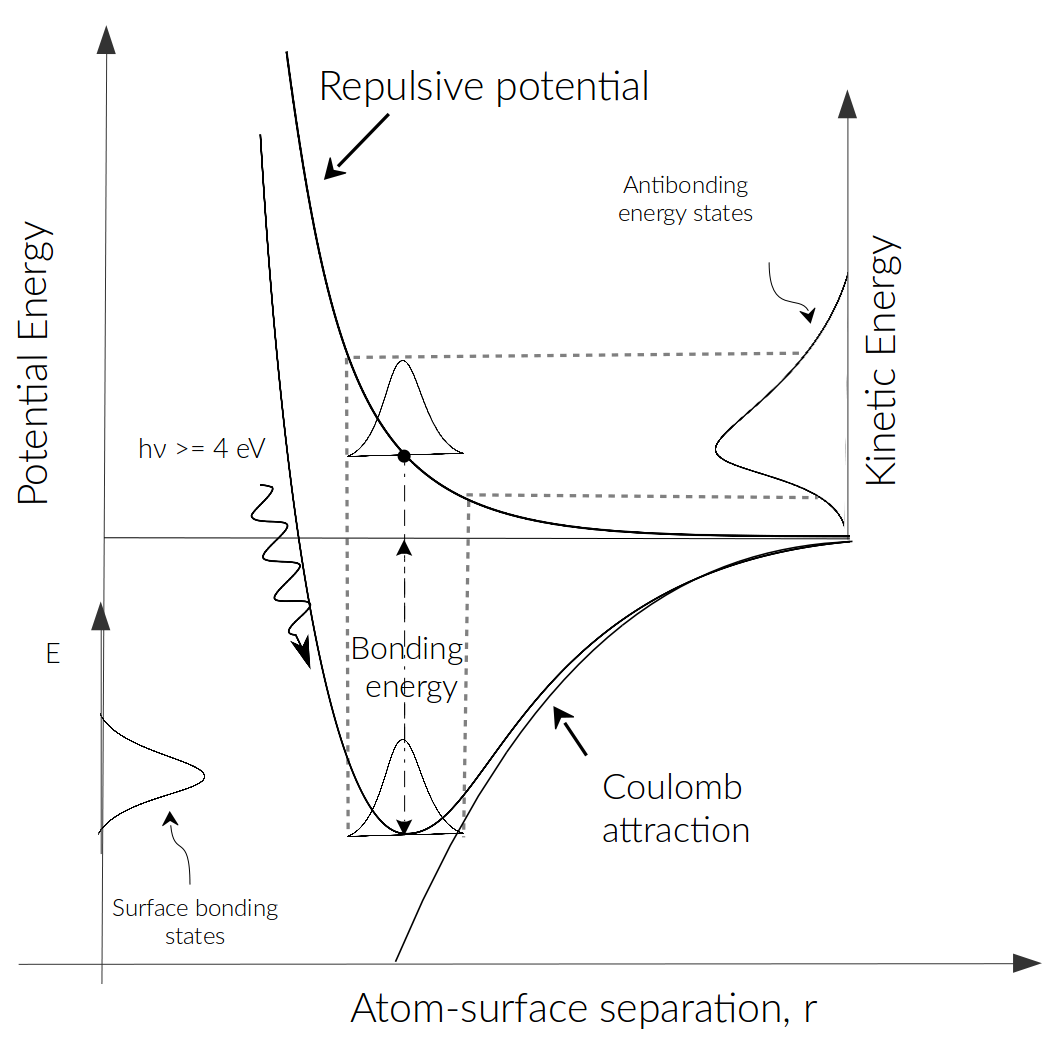}
\caption{Schematics of the potential energy as function of the atom-surface 
separation, $r$, of the photon-stimulated desorption of surface species.}  
\label{poten} 
\end{figure}

Figure \ref{poten} illustrates a typical DIET process where a bond of an atom on 
the surface is excited into an antibonding state induced by electron or photon 
absorption through a valence or core hole ionization process. \cite{Mad68} showed
that valence excitations that include one-electron processes can lead to a long 
lasting (of the order of 10$^{-15}$ to 10$^{-14}$~s) antibonding repulsive state, 
from which desorption occurs. \cite{Mad68} also showed that ``\textit{ESD and PSD 
of ions from both covalently bonded and ionically bonded surface species proceed 
through multielectron excitations that produce highly repulsive electronic states}''; 
these states have sufficiently long lifetimes ($\sim 10^{-14}$~s) that electronic 
energy can be converted to atomic motion, i.e., kinetic energy.

Such electronic excitation events are by nature non-thermal, therefore it is expected 
that the speed EDF/VDF of the species in the anti-bonding state has 
to be non-thermal with a high-energy tail.

%.................................................................................%% 
\section{Experiments}
\label{exp}

\cite{Yak00, Yak04} studied the desorption induced by electronic transitions (DIET) 
of Na adsorbed on amorphous, stoichiometric SiO$_2$ films and on a lunar basalt sample. 
Experiments included ESD and PSD as release processes. Reported measurements 
were done with a different coverage and and different substrate temperature: 
$\sim 0.22$~ML of Na adsorbed at $\sim$250~K on SiO$_2$ films \cite{Yak00}, and 
$\sim 0.5$~ML of Na adsorbed at 100~K on a lunar basalt sample \cite{Yak04}. Details 
of how the experiments were performed can be found in the respective references.

What is important to mention is that their results show that the ESD and PSD of Na from 
SiO$_2$ films and lunar basalt samples occur at threshold photon energies as low as 
h$\nu\sim$4~eV and that desorbing atoms have suprathermal velocities. 
Although they used ESD and PSD as release mechanisms, they only reported the VDF of 
Na released via ESD and later considered ESD results applicable to PSD as they are 
equivalent. 

The VDFs were interpreted by \cite{Yak00} as the ability of photons 
with threshold energies $\geq$3--4~eV to induce electron transfer from an electronic 
state in the bulk, or from a surface state, to the unoccupied Na$^+$ 3s level. 
When the 3s level is occupied, the Na$^0$ atom has a larger radius compared to the 
original Na$^+$ ion thus the atom is now in a highly repulsive configuration, from
which desorption from the surface occurs.

The velocity distribution for desorbing Na$^0$ from SiO$_2$ films at 250~K we use in
this work can be seen in Figure 2.(a) in \cite{Yak00}. Later, 
\cite{Yak04} did similar experiments on a lunar basalt sample 
at 100~K and the reported VDF is shown in Figure~3 in their work. 
The peaks of the VDFs corresponding to the 250~K and 100~K substrate 
temperatures were reported as 1000~m/s and 800~m/s, respectively. The latter was 
interpreted by the same authors as being similar to a $\sim 900$~K Maxwellian distribution. 
This temperature is obtained only when assuming that the peak speed of the distribution 
is equal to the thermal speed. Similarly, a $\sim 1650$~K Maxwellian would be obtained for 
the VDF corresponding to the 250~K substrate, but in the literature 
people have used Maxwellian VDFs in the range of $1200$--$1500$~K.

%.................................................................................%% 
% Section 4
\section{Velocity distribution functions}
\label{prob}

To mathematically best describe the published laboratory measurements \cite{Yak00, Yak04} 
and planetary observations (e.g.\ \cite{Cass15}) we seek a VDF that has a 
characteristic energy significantly higher than what corresponds to the surface 
temperature and that tails towards higher speeds. The second goal of the seeked model 
distribution function is a parametrization that allows for its applications at 
other surface temperatures than the measured ones, in particular for surfaces of 
Mercury and the Moon.

In the following we will present three different VDFs, two of which have been used 
in the literature, namely the Maxwell-Boltzmann distribution and an ad-hoc distribution 
originally used for the icy surfaces of Jupiters moons \cite{John02}. The third 
distribution is a Weibull distribution, which has not been used so far to model the VDFs for 
ESD and PSD particle release. 

Because no angular distribution is given in the experiments, we assume that the setup
was in such a way that the field of view of the detector observing Na atoms that moved 
in just one direction. This leads us to consider only one-dimensional VDF.
Furthermore, for our derivations of the VDFs we assume that the main influencing 
parameter is the surface temperature. The other two parameters that could influence 
the VDF is the Na coverage and the substrate material. Because of the limited data available we can not take these into 
consideration in this work. We discuss the implications of these assumptions in Section~\ref{dis}.

\subsection{Maxwell-Boltzmann distribution function}

The one-dimensional Maxwell-Boltzmann VDF, which reads in its normalised form as:

\begin{equation}
f(v - v_0)= \left( \frac{m}{2\pi k_B T}\right)^{1/2} \phantom{x}\mathrm{exp}\left[-\displaystyle \frac{m}{2k_BT} (v - v_{0})^2\right]
\label{maxb}
\end{equation}

\noindent where $k_B$ is the Boltzmann constant, $T$ is the gas temperature in 
Kelvin, $v_0$ is the speed offset in m/s. 

If we assume that the most probable speed (peak of the distribution) is equal to 
the thermal speed of the particles released from the surface, we can derive 
the characteristic temperature from $T= m\bar{v}^2/(2k_B)$, where $k_B$ is the 
Boltzmann constant. Therefore, a temperature of $T \sim 1400$~K is obtained if the 
peak of the distribution is at $\bar{v} \sim 1000$~m/s, which corresponds to the 
released Na peak speed measured from a 250~K substrate \cite{Yak00}. Similarly, a 
temperature of $T \sim 900$~K is obtained from the peak of the distribution  at 
$\bar{v} \sim 800$~m/s, which is the case of the peak in speed measured by 
\cite{Yak04} from a lunar basalt sample at 100~K. Nevertheless, a  Maxwellian with 
$T=1200$~K has been more commonly assumed for the former measurements, and it is the 
one we will use in this work.

Figure~\ref{speed_pdfs}  shows the measured VDFs and the various fitted 
VDFs. The \textit{grey-diamonds} symbols in all the plots of Figure~\ref{speed_pdfs} 
are the measured VDFs of the released Na: in the left column are 
measurements on a 250~K substrate and in the right column on a 100~K substrate. The
\textit{dashed-black} curves in Figures \ref{speed_pdfs} (a) and (d) are the M-B fits,
with a temperature of $1200$~K and $900$~K, and shifted to higher speed by an 
offset of 1000~m/s and 730~m/s, respectively. 

We searched for a better fit, which is easily obtained if we use a smaller 
temperature of the M-B distribution, which is still higher than the surface temperature. 
Moreover, to obtain a good fit, these M-B distributions had to be shifted to higher speeds. 
These fits are presented as the \textit{solid-black} curves in Figures~\ref{speed_pdfs} 
(a), with a $300$~K M-B, and (d), with a $150$~K M-B; and shifted in speed by an offset 
of 1000~m/s and 750~m/s, respectively.

From here on we will address the $150$~K and the $300$~K M-B distributions as 
``low'' temperature M-B fits, and the $900$~K and the $1200$~K M-B distributions 
as the ``high'' temperature M-B fits.

\subsection{Empirical PSD distribution function}

We present here the empirical energy distribution function, EPSD, for PSD at Mercury 
and the Moon \cite{Wu03, Wu07, Wu10}, $f(E)_{\rm PSD}$. It is based on the distribution 
given by \cite{John02}, and which was adapted for Mercury's surface by accounting for 
the extra energy in the desorption process and by including an energy cut-off. The 
normalised distribution is given as: 

\begin{equation}
f(E)_{\rm PSD} = \beta (1+\beta)\frac{EU^{\beta}}{(E+U)^{2+\beta}}\left( 1 - \sqrt{\frac{E + U}{E_{max}}} \right) 
\label{pdfepsd}
\end{equation}

\noindent where $U$ is the characteristic energy of particles released by PSD, 
$\beta$ is the shape parameter of the distribution, and $E_{max}=4$~eV is the maximum 
energy the released particles can have based on the available energy of the photons for the reported experiments. 
The characteristic energy, $U$, is related to the surface temperature by \cite{Wu03}: 

\begin{equation}
U=\frac{k_BT_0}{e} \phantom{xx} \mathrm{in} \phantom{x} \mathrm{[eV]} \phantom{xx} \mathrm{with} \phantom{xx} T_0= T_i + T_s,
\end{equation}

\noindent where $e$ is the elementary charge, $T_s$ is the local surface 
temperature, and $T_i$ is the energy contribution by the desorption process, which 
is species specific; for sodium we used $T_{\mathrm{Na}}=600$~K. This is based on 
observations at Mercury \cite{Kill99}.

To derive the VDF from the energy distribution stated above, we 
substitute $E$ for $\frac{mv^2}{2e}$ into Eq.~\ref{pdfepsd} to have the appropriate 
units. First, the cut-off term is re-written as:

\[
1 - \sqrt{\frac{E + U}{E_{max}}} =  1 - \sqrt{\frac{\frac{m}{2e}v^2 + U e}{v_{max}}} 
\]

\noindent where $v_{max}=\sqrt{2eE_{max}/m}$. The new normalization constant 
is obtained after integrating in the range $v \in [0,\infty)$, as follows:

\begin{equation}
1 = \bigintsss_0^{\infty} f(E)_{\rm PSD} dE = \bigintsss_0^{\infty} \frac{mv}{e} \cdot f(v)_{\rm PSD} dv 
\end{equation}

\noindent where we have used: $dE=\frac{mv}{e}dv$. Thus, we arrive at the VDF:

\begin{equation}
f(v)_{\rm PSD} = \beta (1+\beta) \frac{\displaystyle m^2/(2e^2)}{\displaystyle(m/2e)^{2+\beta}} \phantom{x} \frac{v^3 U^{\beta} }{(v^2+2eU/m)^{2+\beta}} \left( 1 - \sqrt{\frac{\frac{m}{2e}v^2 + U e}{v_{max}}} \right)
\label{epsd}
\end{equation}

We perform two different fits using Eq.~\ref{epsd}: the first is using the same 
parameters adapted for the release of Na atoms from the surface of Mercury proposed 
by \cite{Wu10}, i.e., $\beta=0.7$ and $T_0 = T_i + T_s$, where $T_{\mathrm{Na}}=600$~K 
and $T_s=[250, 100]$~K. These fits are shown as the \textit{dashed-black} curves in 
Fig.~\ref{speed_pdfs} (b) and (e) and correspond to the $T_0=850$~K and the $T_0=700$~K. 
Secondly, we use the same Eq.~\ref{epsd} but without an offset in temperature, 
same shape parameter, $\beta=0.7$, and an arbitrary offset towards higher 
speeds to obtain a better fit. These fits are shown as the \textit{solid-black} curves, 
with an offset of $v_0 \sim 500$~m/s for the distribution of the 250~K measurements 
and $v_0 \sim 400$~m/s for the distribution of the 100~K measurements; both with the 
same shape parameter $\beta=0.3$.

\subsection{Weibull distribution function}
The Weibull distribution allows for a wide range of shapes using only two parameters for its 
definition. The normalised Weibull distribution for the random variable $v$ is defined as:
 
\begin{equation}
  f(v;\lambda, \kappa) =
  \begin{cases}
        \frac{\kappa}{\lambda}\left( \frac{v}{\lambda} \right)^{\kappa - 1} \displaystyle \mathrm{e}^{-(v/\lambda)^{\kappa}}  & : v \geq 0, \\
       0 & : v < 0
  \end{cases}
\end{equation}

\noindent where $\kappa$ is the dimensionless \textit{shape parameter} and $\lambda>0$ 
is the \textit{scale parameter} of the distribution (in m/s). The scale parameter 
$\lambda$ is obtained after calculating the mean (first central moment) of the 
probability distribution function:

\[
\bar{v}=  \bigintsss_{-\infty}^{\infty}vf(v, \lambda, \kappa) dv = \bigintsss_0^{\infty} v \frac{\kappa}{\lambda} \left(\frac{v}{\lambda} \right)^{\kappa-1} \mathrm{e}^{-(v/\lambda)^{\kappa}} dv = \lambda \Gamma\left(1+\frac{1}{\kappa}\right)
\]

\noindent It follows that: 

\[
\lambda = \displaystyle \frac{\bar{v}}{\Gamma \left( 1+\frac{1}{\kappa} \right)}.
\]

The surface, which is the staring point of the desorbed atoms, has a given 
temperature $T_s$. This surface temperature will cause an energy broadening of the electronic transition 
induced by the adsorption of the UV photon (as depicted in Fig.~\ref{poten}). Therefore, 
the related kinetic energy of the desorbed Na of $\displaystyle \frac{1}{2}m\bar{v}^2=\frac{3}{2}k_BT_s$ 
folds into the distribution. Since we consider the one-dimensional case, we have 
$\bar{v}=\displaystyle \sqrt{\frac{3 k_B T_s}{m}}$. Substituting $\bar{v}$ in the expression for $\lambda$:

\begin{equation}
\lambda = \displaystyle \frac{\sqrt{\frac{3 k_B T}{m}}}{\Gamma \left( 1+\frac{1}{\kappa} \right)}
\label{lambda}
\end{equation}

\noindent thus, the normalised Weibull distribution for $v \geq 0$ is:

\begin{align}
f(v,v_0,\kappa) = &\kappa \phantom{.} \Gamma \left( 1+\frac{1}{\kappa} \right) \left(\frac{m}{3 k_B T_s} \right)^{1/2} \left[ (v - v_0) \phantom{.} \sqrt{\frac{m}{3 k_B T_s}} \Gamma \left( 1+\frac{1}{\kappa} \right) \right]^{\kappa -1} \times \nonumber \\
                       &\times \mathrm{exp} \left[- \left( (v-v_0)  \phantom{.} \sqrt{\frac{m}{3 k_B T_s}} \Gamma \left( 1+\frac{1}{\kappa} \right) \right) ^{\kappa} \right]
\end{align}

\noindent where $v_0$ is the offset speed, and $\kappa$ is the shape parameter, which is an 
implicit function that is usually determined by numerical means (see \cite{Bhat10}). 

The Weibull fits are represented as the \textit{solid-black} curves shown in 
Figures  \ref{speed_pdfs} (c) and (f). For both measured VDFs we looked for 
a good fit with the same shape parameter, $\kappa$. We get reasonably good fits for both data 
sets with a single set of parameters using $\kappa =1.7$ and $v_0 = 575$~m/s, and using actual 
the surface temperature $T_s$ to derive $\lambda$ from Eq.~\ref{lambda}. We suggest to use this
set of parameters ($\kappa, v_0$) for modelling the PSD processes for planetary surfaces, since 
only the surface temperature is needed as input parameter for the distribution function. 

Of course, better fits to the two data sets can be achieved with the Weibull distribution 
function allowing for different parameters for the different fits. For the VDF 
from the substrate at 250~K the Weibull distribution with $T_s$ = 250~K and $\kappa=1.8$ with 
an offset of $v_0=600$~m/s makes the best fit, whereas the Weibull distribution with $T_s$ = 100~K, 
$\kappa=1.8$ and an offset of $v_0=500$~m/s fits best the VDF from the 100~K 
substrate. Unfortunately, these derived parameters do not allow any meaningful extension for 
other surface temperatures. 

%.................................................................................%% 
\subsection{Graphical Residual Analysis (GRA)}
\label{gra}

The measurements and the fit functions as shown in the panels of Figures~\ref{speed_pdfs}
from (a)--(f) are useful for showing the relationship between the data and the 
proposed models; however, it can hide crucial details about the fit function. 
Plotting the residuals can help show these details well, and should 
be used to check the quality of the fit. The Graphical Analysis of the Residuals 
(GRA) is a common and powerful technique to determine if the model needs refinement 
or to verify if the underlying assumptions of the model are met. 
The normal probability plot is a useful residual plot of this model, which we also
include in our analysis. In this section we briefly describe the GRA method.

The GRA entails a basic, though not quantitatively precise, evaluation of the 
differences between the observed values of the dependent variable after fitting 
a model to the data (residuals).  The GRA is a visual examination of the residuals 
to look for obvious deviations from randomness.

Let us denote $y_i$ the measured value (i.e., the probability of a Na atom having 
a speed $v_i$) and $f(v_i)$ the  value from the model distribution. 
Thus, for every $v_i$, the residual of the $i^{th}$ observation is simply defined 
as the distance between the measured value and the theoretical value:

\begin{equation}
r_i = y_i - f(v_i) \phantom{xx} i=1,2,\ldots, n.
\end{equation}

\noindent where $n$ is the number of data points.

The corresponding residuals plots for the different fits are shown in the bottom 
left of each panel (a)--(f) in Figure~\ref{speed_pdfs}. An overly simplified but 
straightforward way to assess the adequacy of the fit is by looking at the general 
tendency of the residuals plot along with the absolute value of the sum of residuals 
(shown in Table \ref{results}). If the overall residual values are close to zero, 
then the absolute value of their sum would also be close to zero, which can be a 
direct evidence that the model is a good fit to the observations. Conversely, if 
the residual values tend to be far away from zero, either below or above zero, then 
the absolute value of the sum will not be close to zero. However, if the 
residuals alternate evenly above and below zero, independently of the 
distance to zero, then the absolute value of the sum of the residuals will still 
be close to zero. Thus we have not enough information to conclude whether the 
model fits the observations. 

One way to obtain more information about any tendencies 
in the residuals, for instance: suspected outliers, skewness to right or left, 
light-tailedness or heavy-tailedness, or mixtures of normal distributions is through 
the \textit{normal probability plot} (\cite{Cham83}) because is a graphical 
technique for assessing whether or not the residuals are approximately normally 
distributed. If the residuals were perfectly normally distributed this would 
indicate that residuals have a random variation and then it would be reasonable to 
conclude that the model is  adequate and differences are statistical. In other words, 
residuals that are perfectly normally distributed can be considered as noise. 

The residuals are plotted against a theoretical normal distribution in such a way that 
the points should form an approximate straight line. The normal probability plot matches 
the quantiles of the residuals to the quantiles of a normal distribution. These plots 
are shown as the \textit{solid-grey} lines at the bottom right corner of each panel 
(a)--(f) in Figure~\ref{speed_pdfs}. Departures from this straight line indicate 
departures from normality. In the next section we show and interpret the results 
obtained. We followed the theory of GRA from \cite{Stat96} and also to interpret the results.

%.................................................................................%% 
\section{Results}
\label{res}

\begin{figure}
\centering
\begin{minipage}[c][19cm][t]{.49\textwidth}
  \centering
  \includegraphics[scale=0.17]{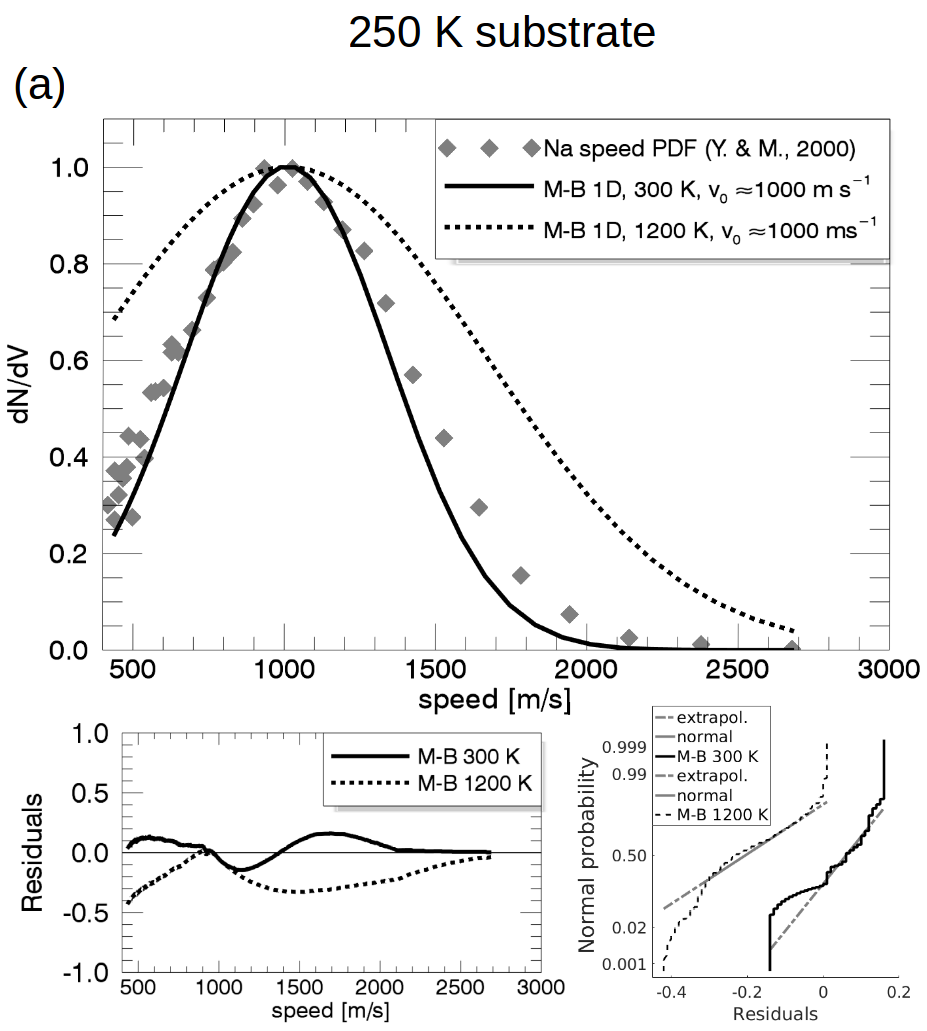}
  \includegraphics[scale=0.17]{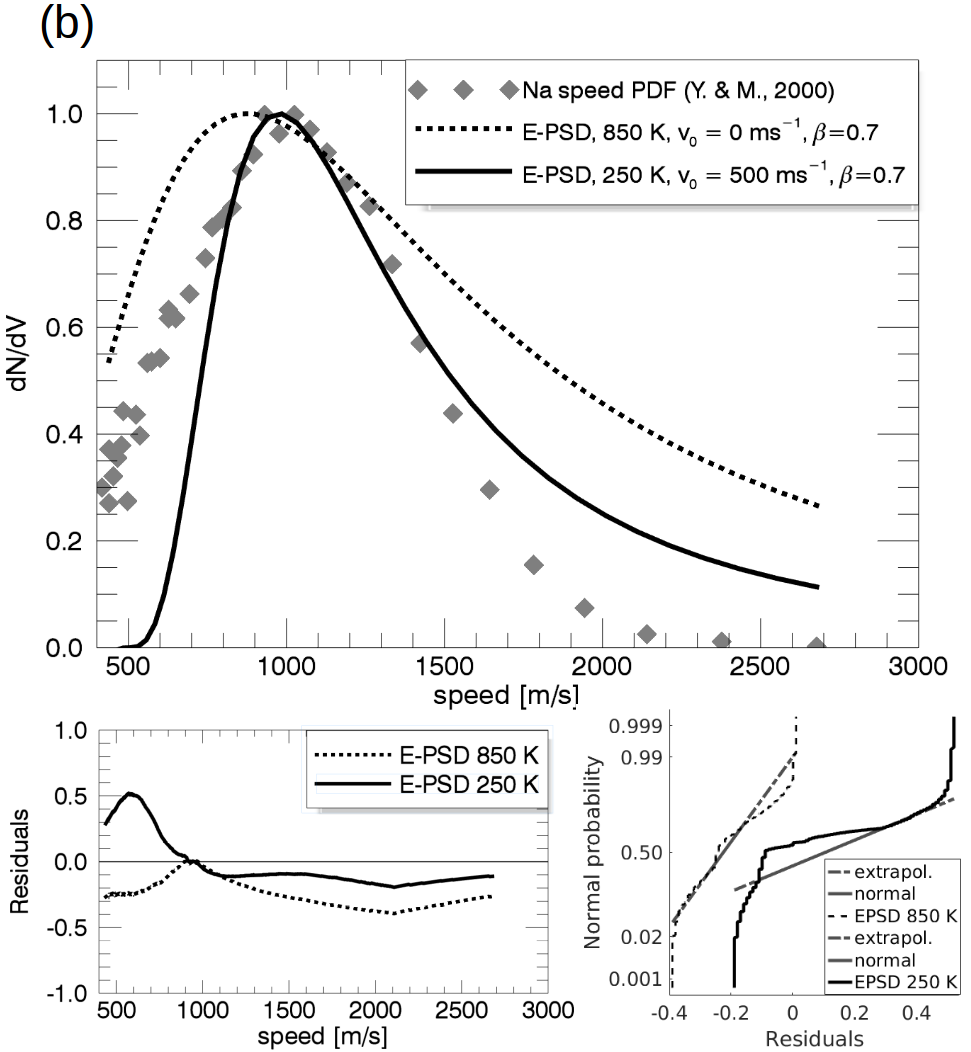}
  \includegraphics[scale=0.17]{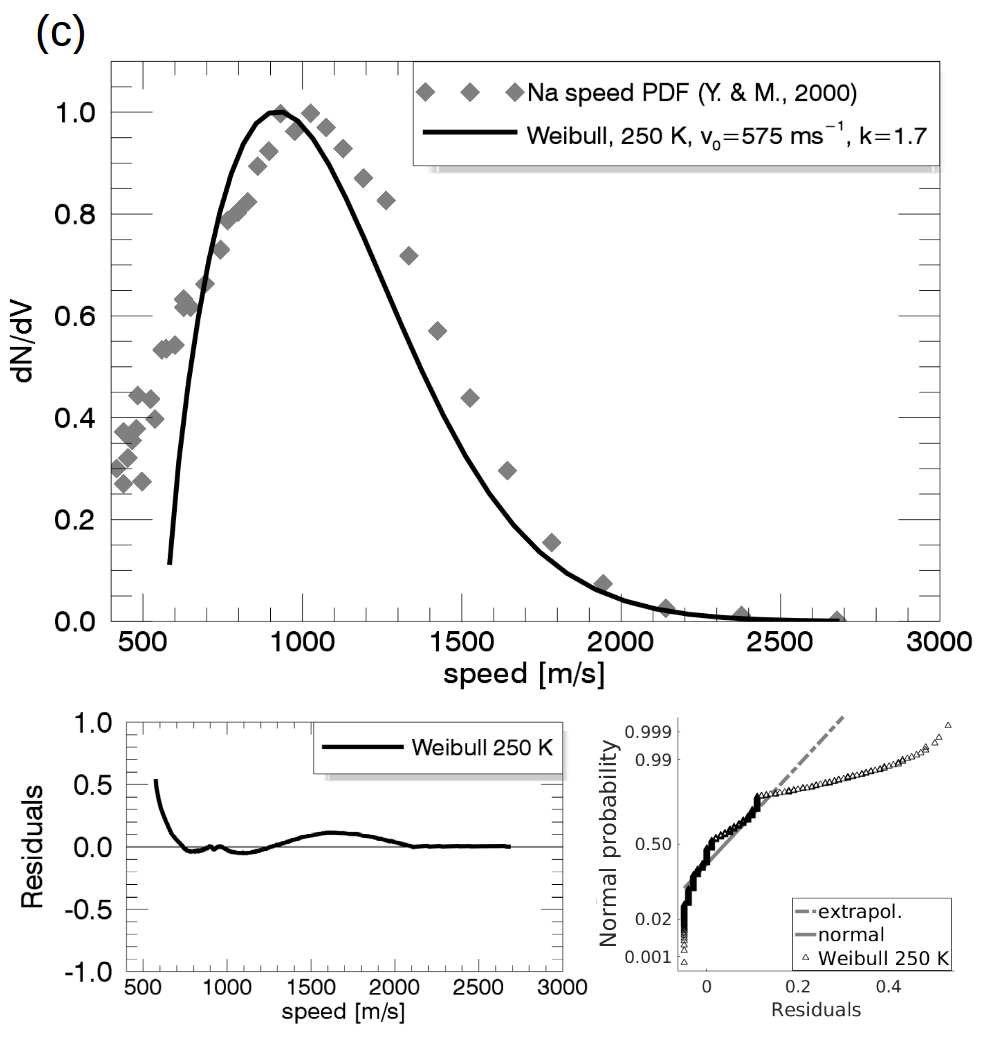}
\end{minipage}
\begin{minipage}[c][19cm][t]{.30\textwidth}
  \centering
  \includegraphics[scale=0.172]{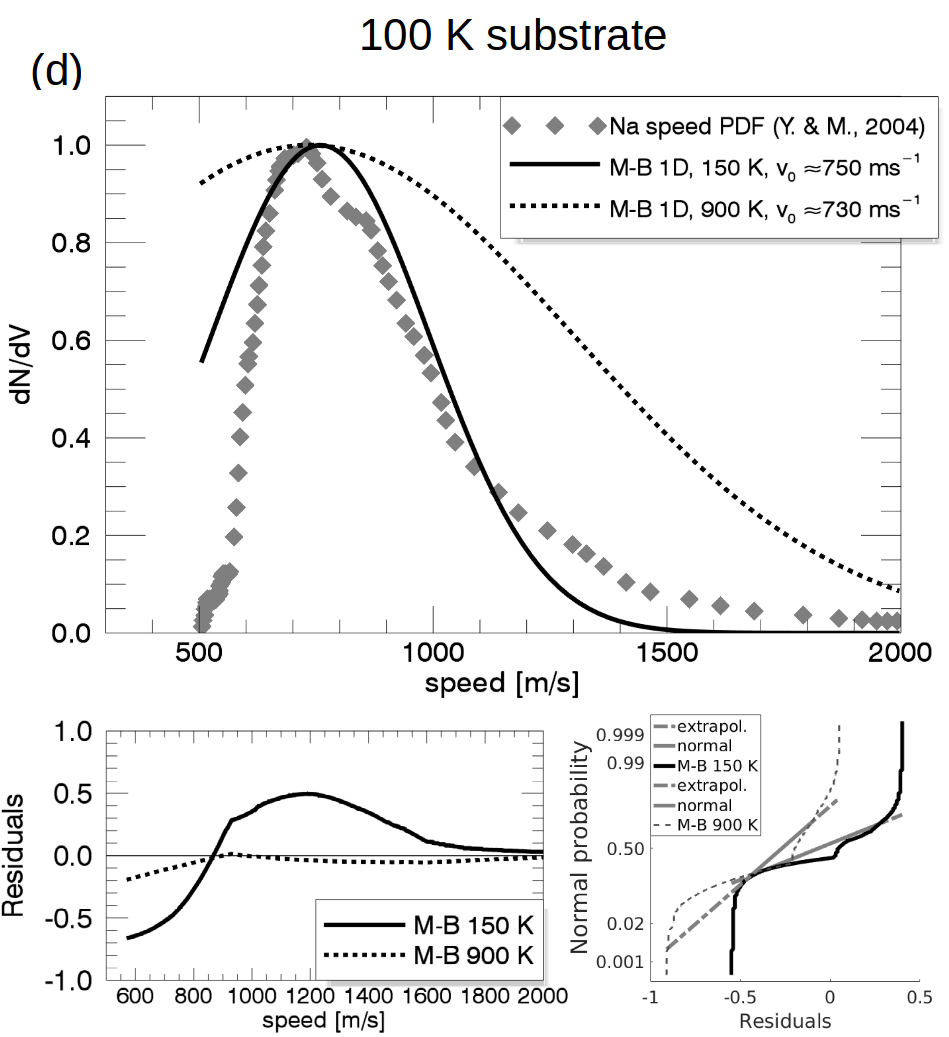}
  \includegraphics[scale=0.17]{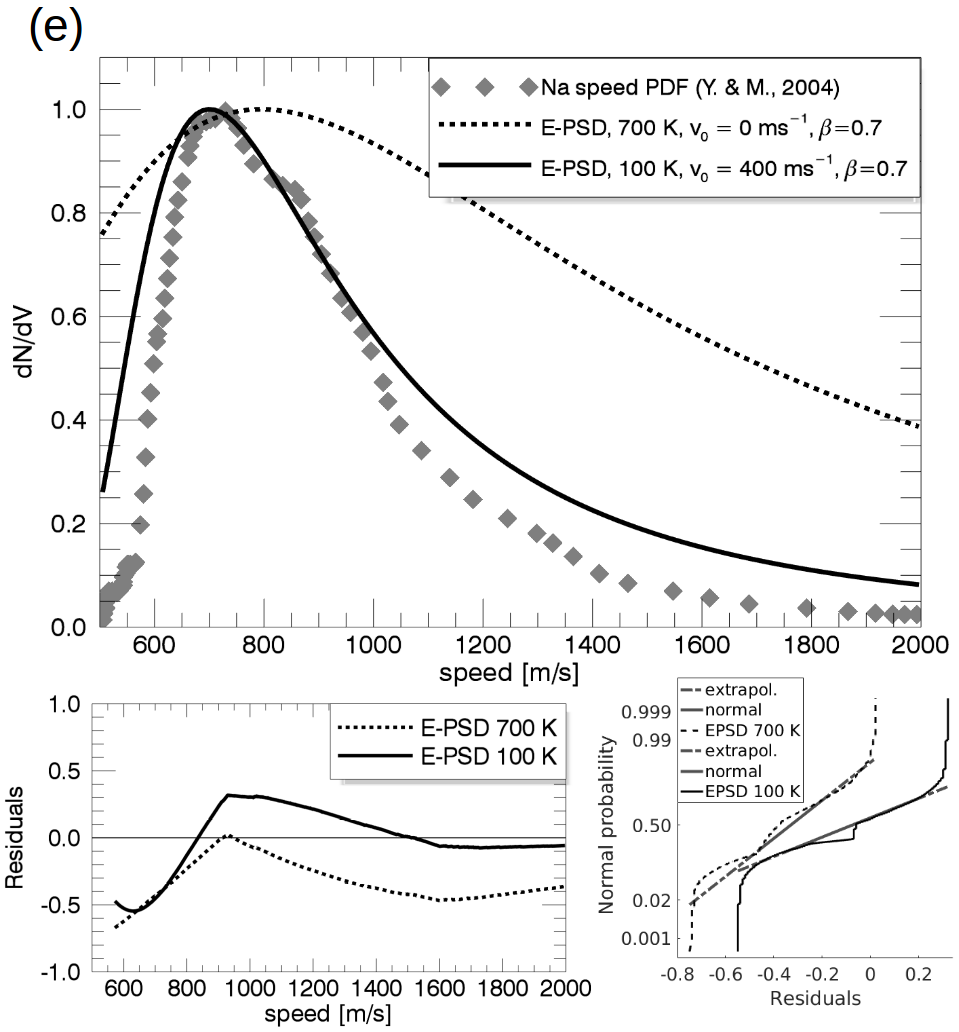}
  \includegraphics[scale=0.17]{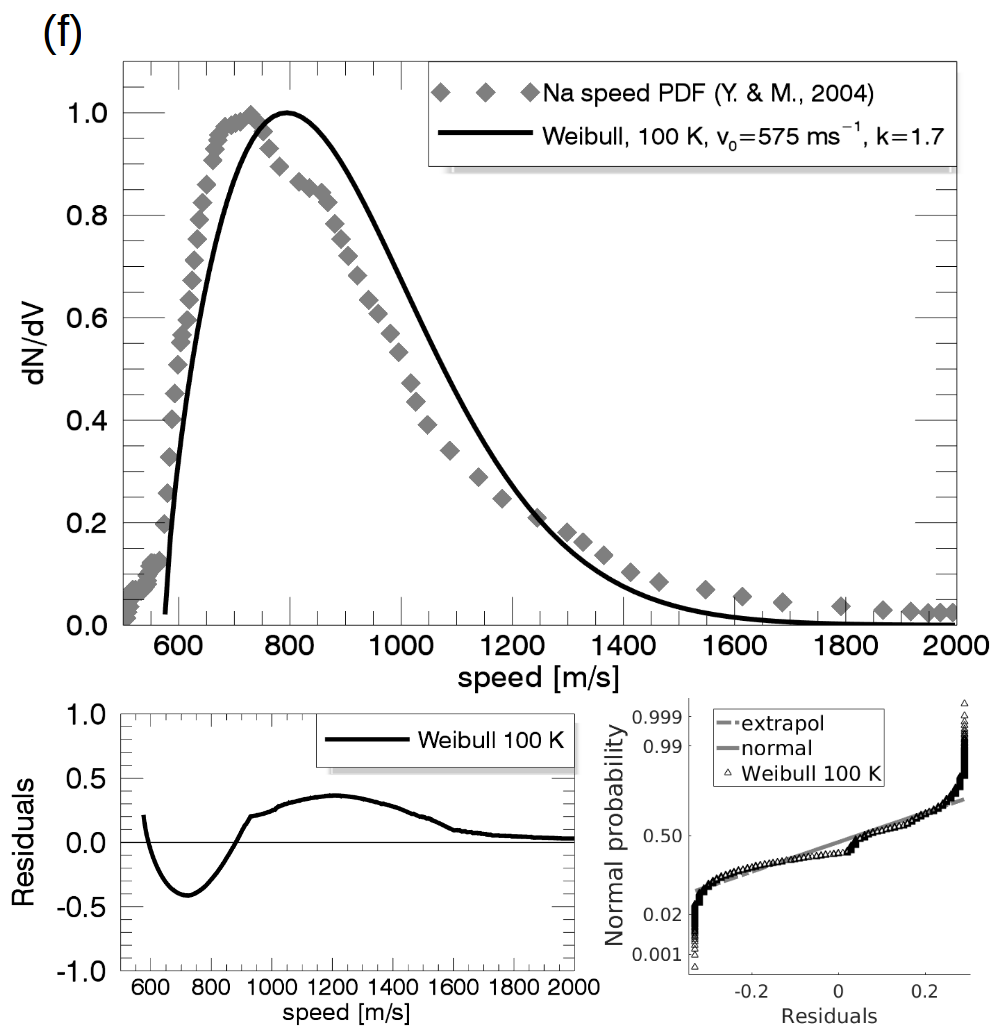}  
\end{minipage}
\caption{Main panels from (a) to (c): \textit{grey-diamonds} are the VDFs from 
neutral Na on a 250~K lunar substrate (see \cite{Yak00}); same from (d) to (f) but
experiments were performed on a lunar sample with a temperature of 100~K (see \cite{Yak04}).
The ``low'' temperature M-B fits are shown as the \textit{solid-black} curves and
the ``high'' temperature M-B as the \textit{dashed-black} curves.  
The \textit{dashed-black} and the \textit{solid-black} curves in (b) and (e)
correspond to the Empirical-PSD distribution proposed by \cite{Wu10}. 
 The \textit{solid-black} curves in (c) and (f) are the Weibull fits. The residuals 
plots and the normal probability plots are shown in the bottom part of the main panels.}
\label{speed_pdfs}
\end{figure}

In Section \ref{prob} we provided the mathematical description of three different 
model distributions and then searched for the best parameter combination to fit the 
measured VDFs under consideration. The different fits obtained are 
presented in this section in the main plots in Figure \ref{speed_pdfs}. How well 
 the model distributions used so far fit the measurements is qualitatively 
estimated by virtue of the GRA. The results of the GRA are shown in the bottom part 
of each panel in Figure~\ref{speed_pdfs}; the plots of residuals  are shown in the 
bottom left and the normal probability plots are shown in the bottom right. 

It is evident from the general features of the Maxwell-Boltzmann fits that the 
``high'' temperature distributions shown in Figure~\ref{speed_pdfs} 
(a) and (d) are much too wide compared to the measured VDFs. In contrast, 
the two ``low'' temperature M-B distributions fit the bulk of the distributions
significantly better, though underestimating the tails, and with the need for large speed 
offsets. 

It is also clear that the E-PSD fits with the parameters adapted for the release
of Na from the surface of Mercury are too wide and shifted compared to the measured 
distributions. Slightly better fits with the E-PSD function are obtained when 
$T_0=T_s$ is used and a speed offset is considered. However, these ``low temperature'' 
fits exhibit very long tails thus overestimating the observed values at higher 
speeds. Although there is no need to adjust an offset temperature, an arbitrary 
offset towards higher speeds is required to fit reasonably well the bulk of the 
measured distributions.

A more careful evaluation of the fits is done with the residuals plots together 
with the normal probability plots. The residuals are plotted against the speed and 
both axes have units of m/s. A reference line at $y=0$ is shown in grey scale to 
identify a change of sign toward positive or negative values.

The residuals of the two ``high'' temperature M-B are mostly 
negative except close to the peak of both distributions where they are almost 
zero. This is because this model function is far above the measured data and 
too wide compared to the measured distribution, which makes the residuals mainly 
have negative values. In fact, the absolute values of the total sum of the residuals 
of these fits are one order of magnitude larger than then rest of the fits, as can 
be seen in Table~\ref{results}. The normal probability plot of the $1200$~K M-B 
distribution shows a strong linear dependence with the theoretical normal 
distribution between the first and the third quartile but it is not centered at 
zero and in the left and right ends the plot bends away from the hypothetical 
straight line. The normal probability plot of the 900~K M-B does not show any 
linear dependence with the theoretical normal distribution and it is not 
symmetrical around zero.

The residuals of the ``low'' temperature M-B fits seem more evenly distributed 
around zero, which makes the absolute value of the total sum of residuals close 
to zero. The normal probability plot of the $300$~K M-B fit exhibits oscillations 
around the straight line between the first and the third quartile and it bends 
above the straight line at the right and left ends. The normal probability plot 
of the 150~K M-B fit has a similar shape compared to the one at 900~K, but the 
plot is well centered around zero.

The residuals of the 850~K and the 700~K E-PSD fits are mostly negative 
but vary with a similar amplitude compared to the residuals of the 250~K and the 
100~K E-PSD fits, which is expected from the shift in speed that the latter 
exhibits with respect to the measured distribution. The normal probability 
plots of the four cases with the E-PSD display a big deviation at the far 
ends and a moderately linear relationship with the theoretical values from 
the standard normal distribution between the first and the third quantile. 
This is a consequence of the heavy-tailed feature of the E-PSD distribution.

The large bend below the straight line displayed in the normal probability plot for 
the Weibull distribution with at temperature of 250~K reveals that the data are 
skewed to the left in comparison with the model distribution. This difference is 
more evident in the main plot between $\sim$450~m/s and $\sim$700~m/s, where the 
Weibull distribution underestimates the observations.

A common aspect of all the normal probability plots obtained is that they all 
exhibit deviations from the normal distribution, specially on either end of the 
speed range. In this case, the parent distribution from which the data were 
sampled is considered to be \textit{heavy-tailed} because the right upper end 
of the normal probability plot bends over the hypothetical straight line that 
passes through the main body of the X-Y values of the normal probability plot, 
and when the left lower end bends below it (\cite{Stat96}).

It is clear from the normal probability plots presented here that none of the 
residuals of the proposed model distributions are normally distributed, i.e., 
that an optimal fit to the data has not been achieved. Nevertheless, it is 
undeniable that the deviations from normality and the absolute value of the 
sum of residuals of the ``high'' temperature M-B are much larger than those 
from the other model distributions. In comparison, the residual are smallest for 
the adapted E-PSD distributions, and Weibull second smallest for the 250~K substrate.

\begin{table}[h]
\centering
\caption{Sum of residuals.}
\label{results}
\begin{tabular}{l l l}
 \hline	
 \textbf{Substrate temperature: 250~K} & Modeled temperature (K) & $ \left| \displaystyle \sum_{i=0}^{n} r_i \right|$\\  
 \hline
	Maxwell-Boltzmann 1D (``low'' temperature) 			& 300		& 49.44	 \\
	Maxwell-Boltzmann 1D (``high'' temperature)			& 1200		& 216.63 \\
	Empirical-PSD										& 250		& 71.80 \\
	Empirical-PSD										& 850		& 234.18 \\
	Weibull												& 250		& 50.49	 \\
	\hline
\textbf{Substrate temperature: 100~K} &  &  \\
  \hline	
	Maxwell-Boltzmann 1D (``low'' temperature) 		& 150			& 48.45  	\\
	Maxwell-Boltzmann 1D (``high'' temperature)		& 900			& 505.04	\\ 
	Empirical-PSD									& 100			& 60.61 	\\
	Empirical-PSD									& 700			& 513.74 	\\	
	Weibull 		  								& 100			& 74.71 	\\
  \hline
\end{tabular} \\
\end{table}

%--------------------------------------------------------------------------------------------
% Section 6
\section{Discussion}
\label{dis}

In this work we test the statistical adequacy of three model distributions: the 
Maxwell-Boltzmann and two non-Maxwellian by means of the Graphical Residual Analysis. 
In this section we discuss which one we consider is physically more valid.

Concerning the results for the M-B fits: we find that the measured VDFs of released 
Na atoms are too narrow compared to the M-B fits suggested by \cite{Yak04} and as 
noted by other authors before. A considerably better fit with M-B is only achieved 
with an offset of the whole distribution to higher speeds and with a lower 
temperature. In all cases of fitting M-B distribution there is no correlation of 
the fit temperature and offset speed with the substrate temperature found. 

Moreover, the applicability of the M-B distribution is limited to gases in 
thermodynamic equilibrium where the energy transfer is facilitated by particle 
collisions, and temperature is interpreted as the mean kinetic energy 
transferred from particle to particle when equilibrium is reached after sufficient 
collisions. These conditions certainly do not apply to the release of atoms by PSD, 
neither on Mercury surface (gas pressure at the surface $p<5\times 10^{-12}$~mbar) 
nor in the experiments done by \cite{Yak00, Yak04} where the residual gas is in 
a non-collisional regime (base pressure of $p < 10^{-10}$~mbar). 
In PSD (and ESD) the energy transfer is carried by single electronic transitions 
on the surface of the substrate. Additionally, the 1D M-B distribution has a symmetric 
shape around the peak but the measured distributions do not exhibit this symmetry. 
The GRA results confirm this, particularly the two ``high'' temperature M-B are 
found to be statistically not  appropriate. This is why the need of a non-thermal 
long-tailed VDF to explain the observed speeds is crucial.

With respect to the results obtained for the non-thermal distributions, 
it is evident that the E-PSD distribution adapted for Na released from the 
Mercury's surface does not provide a good fit to the data, unless we consider 
a different temperature, and we include a rather arbitrary speed offset. This 
increases the amount of free parameters, which is unwanted, because it prevents 
generalisation and extrapolation to other surfaces temperatures.

On the other hand, even though we have no more data to constrain the 
parameters, the Weibull distribution represents the best of the three candidates. 
In terms of the shape, we found that if we use a value of $\kappa=1.7$, we obtain 
the best fit compared to the other distributions, specially the tail of the 
distribution. In terms of statistical adequacy, the GRA results show that the 
residuals are reasonably normally distributed and centered around zero, which is 
of advantage. The Weibull distribution is chosen to fulfill two requirements 
resulting from the observations: first observation is that the mean energy of the 
distribution is significantly above the thermal energy of the surface (given by 
experimental results and observations at Mercury and Moon); second observation 
is that the VDF tails towards higher energies (only from experimental results). 
The Weibull distribution satisfies these two conditions with a minimum number 
of free parameters: the characteristic energy parameter, \textit{$\lambda$}, and 
the shape parameter, \textit{$\kappa$}, as discussed above. Nevertheless, we 
admit that there is no rigorous physics argument for the use of the Weibull 
distribution.

With regard to the fitted offset of the distribution to higher speeds required 
for all the fits we do not have a rigorous explanation. Likely, the excess of 
energy the photon provides after the electronic excitation event, i.e., the 
fraction of the energy is spent in overcoming the Na binding energy during the 
photon-desorption and the remaining fraction is increasing the kinetic energy 
of the released atoms. This offset cannot be explained in terms of a bulk speed 
of the gas, as commonly assumed in the Kinetic Theory of Gases or Plasma Physics, 
where the bulk motion of the gas or plasma produces the offset and the temperature 
is explained by the microscopic motion through collisions. This does not apply to 
the experiments analyzed here since there is no collective behaviour.

Furthermore, and as mentioned in Section \ref{prob}, we do not consider 
as fundamental parameters the Na coverage and the substrate material in our VDF 
derivations. With respect to how the Na coverage affects the VDF, \cite{Yak04} 
found that the ESD/PSD yields scale linearly with Na coverages $< 1$~ML but the 
desorption yield curves are similar \cite{Yak00}. Because of this, we assume that 
there is no substantial difference between a Na coverage of 0.22~ML and 0.5~ML, 
therefore the VDF does not change dramatically either. However, in a similar 
experiment but different substrate material, \cite{Ageev98} found that the ESD peak 
of the energy distributions for Na atoms extend toward low kinetic energies as the Na 
coverage increases above 0.125. Particularly, they found that the low-energy tail 
increases with increasing sodium coverage. In contrast, the results from other
similar experiments by \cite{Mad98, Wil99}, where they study desorption of Potassium, 
which show that when the substrate temperature is kept constant but the K coverage is 
decreased, this leads to a broadening of the VDF and shift of the peak towards 
higher energies, but no low-energy tail is observed when the coverage is increased. 
In any case, it is worth noting that the Na coverage used by \cite{Yak00,Yak04} 
are just experimental; the real coverage of Mercury's or  the Moon's surface is 
likely much lower.

On the other hand, we consider the surface composition of minor importance 
because in \cite{Yak00,Yak04} experiments the Na was applied onto the surface by an 
external dispenser, thus the surface mostly served as a substrate.

Nonetheless, we recognize that the two last assumptions are a 
simplification, particularly given the fact that the lunar sample is a more complex 
oxide compare to the SiO$_2$ films. Unfortunately, there is not enough laboratory 
data to understand the effect of the surface material on the VDF. In this sense, 
the experiments done by \cite{Yak00, Yak04} are merely the starting point for models.

Similar attempts to describe the VDFs reported by \cite{Yak00, Yak04} 
with lower temperatures are, for instance, \cite{Sch13}
who used a Kappa distribution with shape parameter $\kappa$= 1.8, with a 
temperature of 500~K, and with a offset of 0.5 km s$^{-1}$ to match the VDF 
of Na desorbed from a 100~K lunar sample. Whereas they fit a 800~K 
Maxwellian with an offset of 0.2 km s$^{-1}$, providing a good fit to the VDF 
of Na desorbed from a 250~K SiO$_2$ substrate.

%.................................................................................%% 
\section{Conclusions}
\label{concl}

Motivated by an ongoing debate whether or not ESD and PSD produce non-thermal EDF/VDF 
of the desorbed atoms in planetary exospheres, we compare 
the often-used model distributions previously proposed to fit the available observations. 
We use all the available measurements reported by \cite{Yak00, Yak04}, 
who studied the ESD and PSD of Na adsorbed on SiO$_2$ films 
and lunar basalt samples. They reported suprathermal Na atoms with peak speeds of $\sim 800$ and 
$\sim 1000$~m s$^{-1}$, which were  interpreted to come from a 
900~K and a 1200--1500~K Maxwell-Boltzmann distribution, respectively.

Despite that only qualitative support for the non-thermal VDF 
of the released atoms via PSD is available (see Fig. \ref{speed_pdfs}), this study 
helps to confirm that: (1) the Maxwell-Boltzmann distribution is neither statistically nor 
physically adequate to describe non-thermal processes such as ESD and PSD; (2) ESD 
and PSD, being by nature produced by single electronic excitation events, 
produce non-thermal VDFs of the atoms released via these 
processes; and (3) an apparent ``high'' temperature is not needed when a non-thermal 
distribution, such as the Weibull distribution, is considered with the appropriate 
parameters. We recommend to use the Weibull distribution with $\kappa$=1.7, 
$v_0$ = 575 m/s, and the surface temperature to model PSD distributions at planetary 
bodies. 

From observations we know that the large majority of planetary objects of 
the solar system have a surface bounded exosphere and are expected to have 
an extended Na exosphere. It is crucial to choose an appropriate model for the Na atoms 
release from the surface to properly interpret the measurements in planetary 
exospheres. This work is intended to resolve the implications of assuming different
models of atoms released by PSD and ESD from any planetary surfaces not
protected by an atmosphere.

\section*{Acknowledgments}
We thank the Swiss National Science Foundation for supporting this work. 
The data used are listed in the references and included in the supplementary material.

\section*{References}

\bibliography{mybibfile}

\end{document}